%% file: arXiv.tex
\documentclass[aps,prl,twocolumn,showpacs,superscriptaddress,groupedaddress,floatfix]{revtex4-1}  

\usepackage{atlasphysics} 
\usepackage{graphicx}     
\usepackage{subfigure}    
\usepackage{natbib}       
\usepackage{dcolumn}      
\usepackage{bm}           
\usepackage{amssymb}      
\usepackage{amsmath}      
\usepackage{wasysym}      
\usepackage{multirow}     
\usepackage{afterpage}    
\usepackage[colorlinks,hyperindex,breaklinks]{hyperref} 
\hypersetup{linkcolor=blue,citecolor=blue,filecolor=black,urlcolor=blue} 

\usepackage{preprintcover}
\PreprintCoverPaperTitle{Search for Scalar Diphoton Resonances in the Mass Range 65--600~Ge$\kern -0.1em$V with the ATLAS Detector in \textit{\textbf{pp}} Collision Data at \mbox{$\bf \sqrt{s}$ = 8~Te$\kern -0.1em$V}} 
\PreprintIdNumber{CERN-PH-EP-2014-142} 
\PreprintCoverAbstract{A search for scalar particles decaying via narrow resonances into two photons in the mass range 65--600~Ge$\kern -0.1em$V is performed using 20.3~\mbox{fb$^{-1}$} of $\sqrt{s}$ = 8~Te$\kern -0.1em$V \textit{pp} collision data collected with the ATLAS detector at the Large Hadron Collider. The recently discovered Higgs boson is treated as a background. No significant evidence for an additional signal is observed. The results are presented as limits at the 95\% confidence level on the production cross-section of a scalar boson times branching ratio into two photons, in a fiducial volume where the reconstruction efficiency is approximately independent of the event topology. The upper limits set extend over a considerably wider mass range than previous searches.}
\PreprintJournalName{Physical Review Letters}

\begin{document}

\title{Search for Scalar Diphoton Resonances in the Mass Range 65--600~Ge$\kern -0.1em$V with the ATLAS Detector in \textit{\textbf pp} Collision Data at $\bf \sqrt{s}$ = 8~Te$\kern -0.1em$V}

\author{ATLAS Collaboration}

\begin{abstract}
A search for scalar particles decaying via narrow resonances into two photons in the mass range 65--600$\mathrm{\ Ge\kern -0.1em V}$ is performed using 20.3~\mbox{fb$^{-1}$} of $\sqrt{s}$ = 8$\mathrm{\ Te\kern -0.1em V}$ $pp$ collision data collected with the ATLAS detector at the Large Hadron Collider. The recently discovered Higgs boson is treated as a background. No significant evidence for an additional signal is observed. The results are presented as limits at the 95\% confidence level on the production cross-section of a scalar boson times branching ratio into two photons, in a fiducial volume where the reconstruction efficiency is approximately independent of the event topology. The upper limits set extend over a considerably wider mass range than previous searches.
\end{abstract}


\maketitle

In July 2012, the ATLAS and CMS collaborations reported the discovery of a new particle~\cite{Aad:2012tfa,Chatrchyan:2012ufa} whose measured couplings and properties are compatible with the Standard Model Higgs boson ($H$)~\cite{Aad:2013wqa, CMSdiscoverylong, tagkey2013120, Chatrchyan:2012jja}. However, several extensions to the Standard Model, in particular models featuring an extended Higgs sector~\cite{Hill:1987ea,Veltman:1989vw,Binoth:1996au,Schabinger:2005ei,Patt:2006fw,Pruna:2013bma,Lee:1973iz}, predict new scalar resonances below or above the $H$ mass which may be narrow when their branching ratio to two photons is non-negligible.

This Letter presents a search for a scalar particle $X$ of mass \mX\ decaying via narrow resonances into two photons. It extends the method developed for the measurement of the $H$ couplings in the \Hgg\ channel~\cite{Aad:2013wqa} to the range $65 < \mX < 600 ~\GeV$. Analytical descriptions of the signal and background distributions are fitted to the measured diphoton invariant mass spectrum \mgg\ to determine the signal and background yields. The result is presented as a limit on the production cross-section times the branching ratio $BR(X\to\gamgam)$, restricted to a fiducial volume where the reconstruction efficiency is approximately independent of the event topology. The resonance with mass \mX\ is considered narrow when its intrinsic width is smaller than \mbox{$0.09 \GeV + 0.01\cdot\mX$}. This upper limit is defined such that the bias in the number of fitted signal events is kept below 10\%, and ensures that the diphoton invariant mass width is dominated by the experimental resolution in the ATLAS detector. Model-dependent interference effects between the resonance and the continuum diphoton background are not considered.

The ATLAS detector~\cite{Aad:2008zzm} at the LHC~\cite{LHC} covers the pseudorapidity~\footnote{ATLAS uses a right-handed coordinate system with its origin at the nominal interaction point (IP) in the center of the detector, and the $z$-axis along the beam line. The $x$-axis points from the IP to the center of the LHC ring, and the $y$-axis points upwards. Cylindrical coordinates $(r,\phi$) are used in the transverse plane, $\phi$ being the azimuthal angle around the beam line. Observables labeled “transverse” are projected into the $x\text{-}y$ plane. The pseudorapidity is defined in terms of the polar angle $\theta$ as $\eta = \text{--}~\ln [\tan(\theta/2)]$.} range $|\eta|<$ 4.9 and the full azimuthal angle $\phi$. It consists of an inner tracking detector covering the pseudorapidity range $|\eta|< 2.5$, surrounded by electromagnetic and hadronic calorimeters and an external muon spectrometer. 

The search is carried out using the \rts~= 8~\TeV\ $pp$ collision dataset collected in 2012, with stable beam conditions and all ATLAS subsystems operational, which corresponds to an integrated luminosity of \mbox{${\cal L}=20.3 \pm 0.6~\ifb$}~\cite{Aad:2013ucp}. The data were recorded using a diphoton trigger that required two electromagnetic clusters with transverse energies \ET\ above 20 \GeV, both fulfilling identification criteria based on shower shapes in the electromagnetic calorimeter. The efficiency of the diphoton trigger~\cite{ATLAS-CONF-2012-048} is $(98.7 \pm 0.5)\%$ for signal events passing the analysis selection. 

The event selection requires at least one reconstructed primary vertex with two or more tracks with transverse momenta $\pt>0.4~\GeV$, and at least two photon candidates with $\ET > 22~\GeV$ and $|\eta|<2.37$, excluding the barrel/endcap transition region of the calorimeter, $1.37<|\eta|<1.56$.

Photon reconstruction is seeded by clusters of electromagnetic calorimeter cells. Clusters without matching tracks are classified as unconverted photons. Clusters with matched tracks are considered as electron candidates, but are classified as converted photons if they are associated with two tracks consistent with a $\gam\to\ee$ conversion process, or a single track leaving no hit in the innermost layer of the inner tracking detector. The photon energy calibration procedure is the same as in Ref.~\cite{Aad:2013wqa}.

Photon candidates are required to fulfill identification criteria based on shower shapes in the electromagnetic calorimeter, and on energy leakage into the hadronic calorimeter~\cite{ATL-PHYS-PUB-2011-007}. Identification efficiencies, averaged over $\eta$, range from 70\% to above 99\% for the \ET\ range under consideration. To further reduce the background from jets, the calorimeter isolation transverse energy $\ET^{\rm iso}$ is required to be smaller than 6~\GeV, where $\ET^{\rm iso}$ is defined as the sum of transverse energies of the positive-energy topological clusters~\cite{Aad:2010sp} within a cone of size \mbox{$\Delta R=\sqrt{(\Delta\phi)^2+(\Delta\eta)^2} = 0.4$} around the photon candidate. The core of the photon shower is excluded, and $\ET^{\rm iso}$ is corrected for the leakage of the photon shower into the isolation cone. The contributions from the underlying event and pile-up are subtracted using the technique proposed in Ref.~\cite{Salam} and implemented as described in Ref.~\cite{Aad:2011mh}. In addition, the track isolation, defined as the scalar sum of the $\pt$ of the primary vertex tracks with \mbox{$\pt>1~\GeV$} in a \mbox{$\Delta R=0.2$} cone around the photon candidate, excluding the conversion tracks, is required to be smaller than 2.6~\GeV. 
 
The \mgg\ invariant mass is evaluated using the leading photon ($\gam_{1}$) and subleading photon ($\gam_{2}$) energies measured in the calorimeter, the azimuthal angle $\Delta\phi$ and the pseudorapidity $\Delta\eta$ separations between the photons determined from their positions in the calorimeter and the position of the reconstructed diphoton vertex~\cite{Aad:2013wqa}. 

After selection, the data sample consists of a continuum background with dominantly \gamgam, \gam--jet, jet--jet events and Drell--Yan (DY) production of electron pairs where both electrons are misidentified as photons. Two peaking backgrounds arise from the $Z$ boson component of the DY and from $H\to\gamgam$.

To increase the sensitivity, the search is split into two analyses: a categorized low-mass analysis covering the range $65 <\mX<110~\GeV$, and an inclusive high-mass analysis covering $110 <\mX < 600~\GeV$. To provide sidebands on both sides of the tested mass point \mX, the \mgg\ ranges are wider than the \mX\ ranges probed and overlap at the transition between the two analyses. 

The low-mass analysis requires a precise modeling of the DY background, dominated by the $Z$ boson resonance, where both electrons are misidentified as photons, mostly classified as converted photons. The loss of signal sensitivity is mitigated by separating the events into three categories with different signal-to-background ratios, according to the conversion status of the photon pair: two unconverted (UU), one converted and one unconverted (CU) or two converted photons (CC). Table \ref{tab:convcat} shows the fractions of signal and DY events expected in each category. 

\begin{table}[!ht]
    \caption{\label{tab:convcat} Number of diphoton events in data ($N_{\rm data}$), number of expected Drell--Yan events ($N_{\rm DY}$), fractions of expected signal ($f_{X}$) and Drell--Yan ($f_{\rm DY}$) in each conversion category for the low-mass analysis. The signal fraction is given for \mX = 90~\GeV\ but the mass-dependence is negligible. }
    \begin{footnotesize}
    \begin{ruledtabular}
    \begin{tabular}{c|ccc}
        \gam\gam\ category & UU & CU & CC \\
    \hline
        $N_{\rm data}$  & 272184        & 253804         & 63224 \\
        $N_{\rm DY}$   & 1080 $\pm$ 260 & 3400 $\pm$ 600 & 2700 $\pm$ 250\\
    \hline
        $f_{\rm DY}$   & 15.0\%         & 47.3\%         & 37.7\% \\
        $f_{X}$    & 48.7\%         & 42.5\%         & 8.8\% \\
    \end{tabular}
  \end{ruledtabular}
\end{footnotesize}
\end{table}

In each category, the \Zboson\ resonance shape is described by a double-sided Crystal Ball function~\footnote{A double-sided Crystal Ball function is composed of a Gaussian distribution at the core, connected with two power-law distributions describing the lower and upper tails~\cite{Oreglia:1980}.}. Due to the limited size of the fully simulated \Zee\ sample \cite{geant4,atlassim} where both electrons are misidentified as photons, the shape parameters are determined by a fit to a dielectron data sample, where both electrons are required to fulfill shower shape identification criteria and the same \ET\ thresholds as the photons. 

\begin{figure}[!ht]
\includegraphics[width=.49\textwidth]{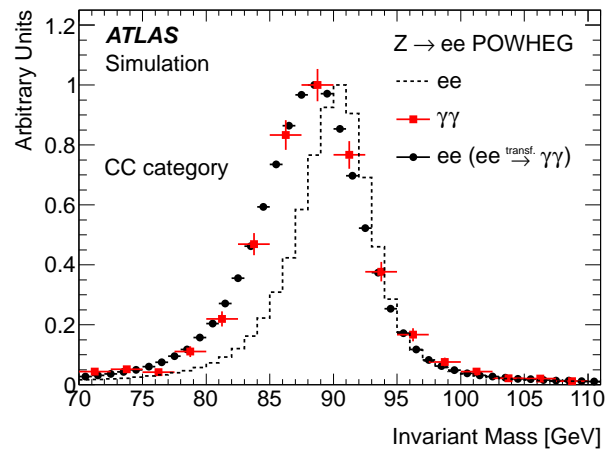}
\caption{Invariant mass distributions in the CC category for fully simulated \Zee\ events reconstructed as $ee$ (dotted-lines), reconstructed as $\gam\gam$ (squares), and reconstructed as $ee$ after transforming the electrons to match the kinematics of the electrons misidentified as converted photons (circles).}
\label{fig:DYtemplates} 
\end{figure}

Since most of the electrons misidentified as photons underwent large bremsstrahlung, the invariant mass distribution of the $Z$ boson reconstructed as a photon pair is wider and shifted to lower masses by up to 2~\GeV\ with respect to the $Z$ boson mass reconstructed as an electron pair. The \Zee\ invariant mass distributions extracted from data in each category are transformed by applying \ET-dependent shifts and smearing factors to the electron \ET\ and $\phi$, to match the kinematics of the electrons misidentified as photons. 
Two sets of transformations are derived for $\gam_{1}$ and $\gam_{2}$ depending on their conversion status, using a \Zee\ sample generated with {\sc powheg}~\cite{Alioli:2008tz, Nason:2009ai} interfaced with {\sc pythia8}~\cite{Sjostrand:2006za} for showering and hadronization. Figure~\ref{fig:DYtemplates} illustrates the effect of the electrons' transformations on the invariant mass shapes in the fully simulated \Zee\ sample. Systematic uncertainties on the template shapes and the $Z$ peak position are evaluated by varying the parameters of the electrons' transformations by $\pm 1\sigma$. 

The DY normalization is computed from the $e\to\gam$ fake rates, defined as the ratios of $e\gam$ to $ee$ pairs measured in \Zee\ data, separately for $\gam_{1}$ and $\gam_{2}$ and each conversion status. A correction factor obtained from fully simulated \Zee\ events is applied to account for additional effects, mainly the differences in isolation efficiencies and vertex reconstruction efficiency between $\gam\gam$ and $ee$ events. The associated uncertainties (9 to 25\%) are dominated by the subtraction of the continuum background and the detector material description.

\begin{figure}[!ht]
\includegraphics[width=.49\textwidth]{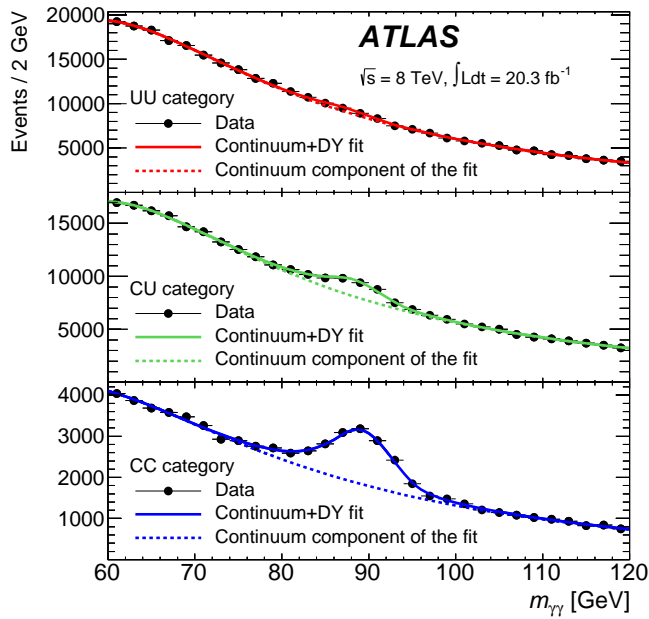}
\caption{Background-only fits to the data (black dots) as functions of the diphoton invariant mass \mgg\ for the three conversion categories in the low-mass range. The solid lines show the sum of the Drell--Yan and the continuum background components. The dashed lines show the continuum background component only. }
\label{fig:backgroundfit_low} 
\end{figure}

The determination of the  analytical form of the continuum background and the corresponding uncertainties follow the method detailed in Ref.~\cite{Aad:2012tfa}. The sum of a Landau distribution and an exponential distribution is used over the full \mgg\ range. The bias on the signal yield induced by the analytical shape function is required to be lower than 20\% of the statistical uncertainty on the fitted signal yield for the background-only spectrum. This bias is measured from a large sample generated from a parameterized detector response, and is accounted for by a mass-dependent uncertainty. Figure \ref{fig:backgroundfit_low} shows background-only fits to the data in the low-mass analysis for the three conversion categories.

In the high-mass analysis, relative cuts \mbox{$\ET^{\gamma_{1}}/\mgg > 0.4$} and \mbox{$\ET^{\gamma_{2}}/\mgg > 0.3$} are added to the selection requirements to reduce the continuum backgrounds and thereby increase the signal sensitivity. In total, 108654 events with \mbox{$100< \mgg <800~\GeV$} are selected. 

To determine the continuum background shape over this large mass range, an exponential of a second-order polynomial is fitted inside a sliding \mgg\ window of width \mbox{$80\cdot(\mX-110~\GeV)/110 + 20 \GeV$}, centered on the mass point \mX. The analytical shape and the fit window width are chosen to fulfill the signal yield bias criterion, as defined for the low-mass analysis, to minimize the statistical uncertainty on the background.

\begin{figure}[!ht]
\includegraphics[width=.49\textwidth]{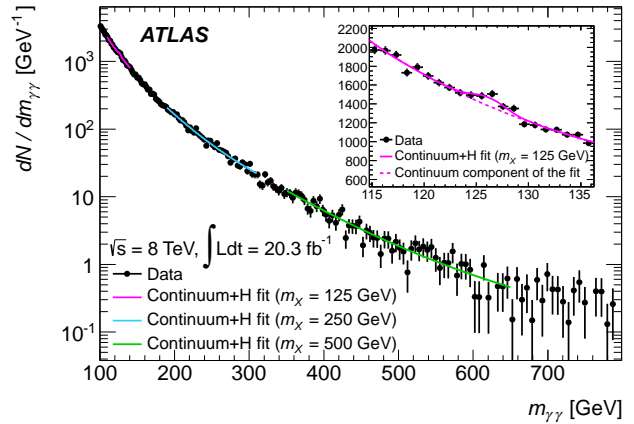}
\caption{Background-only fits to the data (black dots) as functions of the diphoton invariant mass \mgg\ for the inclusive high-mass analysis. The solid line shows the sum of the Higgs boson and the continuum background components. The dashed line shows the continuum background component only. }
\label{fig:backgroundfit_high} 
\end{figure}

The $H$ background shape is modeled by a double-sided Crystal Ball function, and normalized for \mH~=~125.9~\gev~\cite{ParticleDataGroup}\footnote{Differences between this choice of reference mass and the new mass measurements~\cite{ATLASmass,CMSmass} are covered by the energy scale uncertainties listed in Table \ref{tab:systs}.} using the most up-to-date Standard Model cross-section calculations and corrections~\cite{Heinemeyer:2013tqa} of the five main production modes: gluon fusion (ggF), vector-boson fusion (VBF), Higgs-strahlung ($WH, ZH$), and associated production with a top quark pair (\ttbar\Hboson). The ggF and VBF samples~\cite{Aad:2013wqa} are simulated with the {\sc powheg} generator interfaced with {\sc pythia8}. The $WH$, $ZH$ and \ttbar\Hboson\ samples~\cite{Aad:2013wqa} are simulated with {\sc pythia8}. Figure \ref{fig:backgroundfit_high} shows background-only fits to the data in the high-mass analysis.

\begin{figure*}
\includegraphics[width=0.95\textwidth]{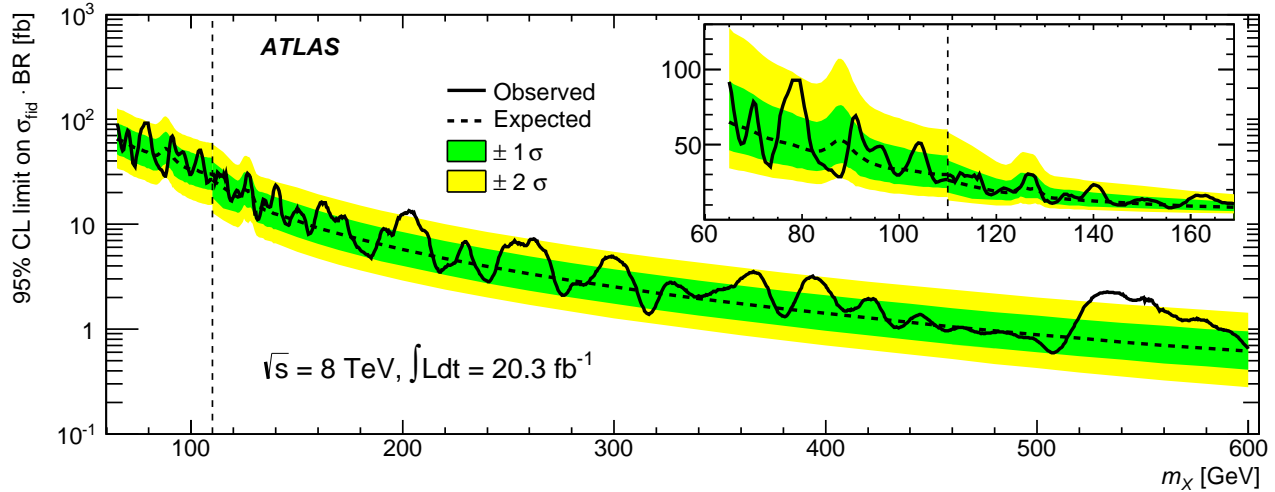}
\caption{Observed and expected 95\% CL limit on the fiducial cross-section times branching ratio BR($X\to\gamma\gamma$) as a function of \mX\ in the range $65 < \mX < 600~\GeV$. The discontinuity in the limit at \mbox{\mX\ = 110 \GeV} (vertical dashed line) is due to the transition between the low-mass and high-mass analyses. The green and yellow bands show the $\pm 1\sigma$ and $\pm 2\sigma$ uncertainties on the expected limit. The inset shows a zoom around the transition point.}\label{fig:limits} 
\end{figure*}

The expected invariant mass distribution of the narrow resonance signal $X$ is also modeled with a double-sided Crystal Ball function in the mass range \mbox{$65 \leq \mX \leq 600~\GeV$}, using fully simulated ggF($X$) samples generated as for $H$, where $H$ is replaced by a scalar boson with a constant width of 4~\MeV. Polynomial parameterizations of the signal shape parameters as a function of \mX\ are obtained from a simultaneous fit to all the generated mass points \mX, separately for the high-mass analysis and the three low-mass analysis categories. The signal shape parameters extracted from ggF($X$) are compared to the other production modes VBF($X$), $\Wboson X$, \Zboson$X$ and \ttbar$X$; the bias on the signal yield due to the choice of ggF($X$) shape is negligible. The systematic uncertainty on the signal shape due to the photon energy resolution uncertainty ranges from 10\% to 40\% as a function of \mX~\cite{Aad:2013wqa}. The systematic uncertainty on the $X$ peak position due to the photon energy scale uncertainty is 0.6\%~\cite{Aad:2013wqa}.

The fiducial cross-section \mbox{$\sigma_{\rm fid} \cdot BR(X\to\gamma\gamma)$} includes an efficiency correction factor $C_{X}$ through
$$\sigma_{\rm fid} \cdot BR(X\to\gamma\gamma) = \frac{N_{\rm data}}{C_{X}\cdot\cal{L}}~~\mathrm{with}~~C_{X}=\frac{N_{\rm MC}^{\rm reco}}{N_{\rm MC}^{\rm fid}},$$
where $N_{\rm data}$ is the number of fitted signal events in data, $N_{\rm MC}^{\rm reco}$ the number of simulated signal events passing the selection criteria and $N_{\rm MC}^{\rm fid}$ the number of simulated signal events generated within the fiducial volume. 
The fiducial volume, defined from geometrical and kinematical constraints at the generated particle level, is optimized to reduce the model-dependence of $C_{X}$ using fully simulated samples of the five $X$ production modes to cover a large variety of topologies. The photon selection at generation level is similar to the selection applied to the data: two photons with $\ET > 22~\GeV$ and $|\eta| < 2.37$ are required; for \mX\ greater than 110 \GeV, the relative cuts \mbox{$\ET^{\gamma_{1}}/\mgg > 0.4$} and \mbox{$\ET^{\gamma_{2}}/\mgg > 0.3$} are imposed. 
The particle isolation, defined as the scalar sum of \pt\ of all the stable particles (except neutrinos) found within a $\Delta R=0.4$ cone around the photon direction, is required to be less than 12 \GeV.
The $C_{X}$ factor is parameterized from the ggF($X$) samples, and ranges from 0.56 to 0.71 as a function of \mX. Systematic uncertainties include the maximum difference between the $C_{X}$ of the five production modes, the effect of the underlying event (U.E.) and pile-up. 

The statistical analysis of the data uses unbinned maximum likelihood fits. The DY and H shapes and normalizations are allowed to float within the uncertainties. In the low-mass analysis, a simultaneous fit to the three conversion categories is performed. Only two excesses with 2.1$\sigma$ and 2.2$\sigma$ local significances above the background are observed over the full mass range \mbox{65--600 \GeV}, for \mbox{\mX=201 \GeV} and \mbox{\mX=530 \GeV} respectivelly. This corresponds to a deviation of less than 0.5$\sigma$ from the background-only hypothesis. Consequently, a $95\%$ limit on $\sigma_{\rm fid} \cdot BR(X\to\gamma\gamma)$ is computed using the procedure of Ref.~\cite{Aad:2012tfa}. The systematic uncertainties listed in Table \ref{tab:systs} are accounted for by nuisance parameters in the likelihood function. In the low-mass analysis, the dominant uncertainties are the DY normalization and the residual topology dependence of $C_{X}$. In the high-mass analysis, the largest uncertainties arise from the energy resolution and the theoretical uncertainty on the production rate of the Standard Model Higgs boson around 126~\GeV. 

The observed and expected limits, shown in Fig.~\ref{fig:limits}, are in good agreement, consistent with the absence of a signal. The limits on $\sigma_{\rm fid} \cdot BR(X\to\gamma\gamma)$ for an additional scalar resonance range from 90~fb for \mbox{\mX = 65~\GeV} to 1~fb for \mbox{\mX = 600~\GeV}. These results extend over a considerably wider mass range than the previous searches by the ATLAS and CMS collaborations~\cite{Aad:2012tfa,CMSHgamgamFinal}, are complementary to spin-2 particles searches~\cite{ED_gamgam_ATLAS,ED_gamgam_CMS}, and are the first such limits independent of the event topology.

\begin{table}
\caption{\label{tab:systs} Summary of the systematic uncertainties}
\begin{footnotesize}
\begin{ruledtabular}
\begin{tabular}{ll|ll}
  \multicolumn{2}{l|}{\em Signal and Higgs boson yield}   & \multicolumn{2}{l}{\em Z component of Drell--Yan} \\
      \hspace{.2cm} Luminosity  & $2.8$\% & \hspace{.2cm} Normalization\footnotemark[2]  & $9$--$25$\%\\
      \hspace{.2cm} Trigger     & $0.5$\% &  \hspace{.2cm} Peak position\footnotemark[2]  & $1.5$--$3.5$\%\\
      \hspace{.2cm} \gam\ identification\footnotemark[1] & $1.6$--$2.7$\% & \hspace{.2cm} Template shape\footnotemark[2] & $1.5$--$3$\% \\
      \hspace{.2cm} \gam\ isolation\footnotemark[1]   & $1$--$6$\% & \multicolumn{2}{l}{\em Higgs boson background} \\
      \hspace{.2cm} Energy resolution\footnotemark[1]\footnotemark[2]&  $10$--$40$\%  & \hspace{.2cm} Cross-section\footnotemark[3]   & $9.6$\% \\
      \multicolumn{2}{l|}{\em Signal and Higgs boson peak position}  & \hspace{.2cm} Branching ratio & $4.8$\% \\
      \hspace{.2cm} Energy scale   & $0.6$\% & \multicolumn{2}{l}{\em $C_X$ factor}\\
      \multicolumn{2}{l|}{\em Continuum \gamgam, \gam j, jj, DY} & \hspace{.2cm} Topology\footnotemark[1] & $3$--$15$\% \\
      \hspace{.2cm} Signal bias\footnotemark[1] &  $1$--$67$ events & \hspace{.2cm} Pile-up \& U. E.\footnotemark[1]  & $1.4$--$3.2$\% \\

\end{tabular}
\end{ruledtabular}
\footnotetext[1]{\footnotesize mass-dependent.}
\footnotetext[2]{\footnotesize category-dependent.}
\footnotetext[3]{\footnotesize factorization scale + PDF uncertainties~\cite{Heinemeyer:2013tqa}.}
\end{footnotesize}
\end{table}

\vspace{\baselineskip}
We thank CERN for the very successful operation of the LHC, as well as the
support staff from our institutions without whom ATLAS could not be
operated efficiently.

We acknowledge the support of ANPCyT, Argentina; YerPhI, Armenia; ARC,
Australia; BMWF and FWF, Austria; ANAS, Azerbaijan; SSTC, Belarus; CNPq and FAPESP,
Brazil; NSERC, NRC and CFI, Canada; CERN; CONICYT, Chile; CAS, MOST and NSFC,
China; COLCIENCIAS, Colombia; MSMT CR, MPO CR and VSC CR, Czech Republic;
DNRF, DNSRC and Lundbeck Foundation, Denmark; EPLANET, ERC and NSRF, European Union;
IN2P3-CNRS, CEA-DSM/IRFU, France; GNSF, Georgia; BMBF, DFG, HGF, MPG and AvH
Foundation, Germany; GSRT and NSRF, Greece; ISF, MINERVA, GIF, I-CORE and Benoziyo Center,
Israel; INFN, Italy; MEXT and JSPS, Japan; CNRST, Morocco; FOM and NWO,
Netherlands; BRF and RCN, Norway; MNiSW and NCN, Poland; GRICES and FCT, 
Portugal; MNE/IFA, Romania; MES of Russia and ROSATOM, Russian Federation; JINR; MSTD,
Serbia; MSSR, Slovakia; ARRS and MIZ\v{S}, Slovenia; DST/NRF, South Africa;
MINECO, Spain; SRC and Wallenberg Foundation, Sweden; SER, SNSF and Cantons of
Bern and Geneva, Switzerland; NSC, Taiwan; TAEK, Turkey; STFC, the Royal
Society and Leverhulme Trust, United Kingdom; DOE and NSF, United States of
America.

The crucial computing support from all WLCG partners is acknowledged
gratefully, in particular from CERN and the ATLAS Tier-1 facilities at
TRIUMF (Canada), NDGF (Denmark, Norway, Sweden), CC-IN2P3 (France),
KIT/GridKA (Germany), INFN-CNAF (Italy), NL-T1 (Netherlands), PIC (Spain),
ASGC (Taiwan), RAL (UK) and BNL (USA) and in the Tier-2 facilities
worldwide.

\providecommand{\href}[2]{#2}\begingroup\raggedright\endgroup

\onecolumngrid

\include{atlas_authlist}

\end{document}

%% file: atlas_authlist.tex
\begin{flushleft}
{\Large The ATLAS Collaboration}

\bigskip

G.~Aad$^{\rm 84}$,
B.~Abbott$^{\rm 112}$,
J.~Abdallah$^{\rm 150}$,
S.~Abdel~Khalek$^{\rm 116}$,
O.~Abdinov$^{\rm 11}$,
R.~Aben$^{\rm 106}$,
B.~Abi$^{\rm 113}$,
O.S.~AbouZeid$^{\rm 157}$,
H.~Abramowicz$^{\rm 152}$,
H.~Abreu$^{\rm 151}$
% ...
\bigskip
\\
% ...
$^{11}$ Institute of Physics, Azerbaijan Academy of Sciences, Baku, Azerbaijan\\
% ...
$^{84}$ CPPM, Aix-Marseille Universit{\'e} and CNRS/IN2P3, Marseille, France\\
% ...
$^{106}$ Nikhef National Institute for Subatomic Physics and University of Amsterdam, Amsterdam, Netherlands\\
% ...
$^{112}$ Homer L. Dodge Department of Physics and Astronomy, University of Oklahoma, Norman OK, United States of America\\
$^{113}$ Department of Physics, Oklahoma State University, Stillwater OK, United States of America\\
% ...
$^{116}$ LAL, Universit{\'e} Paris-Sud and CNRS/IN2P3, Orsay, France\\
% ...
$^{150}$ Institute of Physics, Academia Sinica, Taipei, Taiwan\\
$^{151}$ Department of Physics, Technion: Israel Institute of Technology, Haifa, Israel\\
$^{152}$ Raymond and Beverly Sackler School of Physics and Astronomy, Tel Aviv University, Tel Aviv, Israel\\
% ...
$^{157}$ Department of Physics, University of Toronto, Toronto ON, Canada\\
% ...
\end{flushleft}